\documentstyle[11pt,aaspp4]{article}

\def\et{et al.}
\def\kms{km s$^{-1}$}

\def\rhalf{R$_{0.5}$}

\def\solar{\ifmmode_{\mof d\odot}\;\else$_{\mathord\odot}\;$\fi}

\begin{document}

\title{The Star Clusters in the Irregular Galaxy 
NGC 4449\footnote{\rm Based in part
on observations with
the NASA/ESA  Hubble Space Telescope, obtained at the Space Telescope
Science Institute, which is  operated  by AURA, Inc.,
under NASA contract NAS 5-26555}
} 

\author{Andrea E.\ Gelatt\footnote{\rm Current address:  P.O. Box 05-74, 
Grinnell College, Grinnell, IA, 50112 (gelatt@grinnell.edu)}}
\affil{Lowell Observatory, 1400 West Mars Hill Road, Flagstaff, Arizona 86001
USA;
\\gelatt@lowell.edu}

\author{Deidre A.\ Hunter}
\affil{Lowell Observatory, 1400 West Mars Hill Road, Flagstaff, Arizona 86001
USA;
\\dah@lowell.edu}

\and

\author{J.\ S.\ Gallagher}
\affil{Washburn Observatory, University of Wisconsin, 475 N.\ Charter St.,
Madison, Wisconsin 53706 USA;
\\jsg@astro.wisc.edu}

\begin{abstract}

We examine the star clusters in the luminous irregular galaxy NGC 4449.
We use a near-infrared spectrum and broad-band images taken with the
{\it Hubble Space Telescope} to place a limit of 8--15 Myrs on the
age of the bright central object in NGC 4449. Its luminosity and size
suggest that it is comparable to young super star clusters.
However, there is a peculiar nucleated-bar
structure at the center of this star cluster, and we suggest
that this structure is debris from the interaction that has produced
the counter-rotating gas systems and extended gas streamers in the galaxy.

From the images we identify 60 other candidate compact star clusters
in NGC 4449. Fourteen of these could be background elliptical
galaxies or old globular star clusters. 
Of the star clusters, three, in addition to the central object,
are potentially super star clusters, and many others are comparable
to the populous clusters found in the LMC. The star clusters span a large range
in ages with no obvious peak in cluster formation that might be attributed to
the interaction that the galaxy has experienced.

\end{abstract}
 

\section{Introduction}

NGC 4449 is a luminous and actively star-forming
barred Magellanic-type irregular galaxy.
It has an integrated M$_{B}$ of $-$18
and a star formation rate
about twice that of the LMC.
It has been well known for decades that NGC 4449 is unusual
in having neutral hydrogen 
extending to six times its Holmberg radius
(van Woerden, Bosma, \& Mebold 1975;
Bajaja, Huchtmeier, \& Klein 1994).
In addition the HI maps had revealed counter-rotating gas systems:
The gas within a 2\arcmin\ radius of the center is rotating in the
opposite direction to the gas outside that radius.
Recently, Hunter \et\ (1998) reobserved the extended HI gas around
NGC 4449 at higher resolution and
resolved the extended gas into
enormous streamers that wrap around the galaxy.

The enormous streamers 
suggest that NGC 4449 has undergone some type of interaction with
another galaxy (Hunter \et\ 1998, Theis 1999).
A perturber could have passed
by, with or without giving up mass to NGC 4449, and set up
tidal tails and bridges or a perturber
could have passed through the disk.
Furthermore, the presence of counter-rotating systems is often interpreted in
other galaxies as the signature of acquisition of external material
as in a merger (Hernquist \& Barnes 1991; Quinn \& Binney 1992;
Braun \et\ 1994), so a small partner could have been absorbed into
NGC 4449.
In some scenarios the effects of the
encounter could remain for timescales on the order of a Gyr, allowing
the perturber to move out of the neighborhood 
or to have been
digested by NGC 4449.
NGC 4449 is a member of the CVnI ``loose'' cloud of galaxies
and the small irregular DDO 125 is
only 41 kpc away in the plane of the sky.
So, there exist potential interaction partners.

In this paper we examine the nature of the system of star clusters in
NGC 4449 and its clues to the galaxy's evolution.  
The type of clusters, their ages, and distribution could potentially
be affected by interaction with another galaxy.
For example,
if the clusters were all the 
same age, this might suggest that some short violent event, such as the 
interaction of galaxies, caused the formation of the clusters.

Additionally, if the galaxy contains a population of the relatively
unusual super star clusters, we might argue that these clusters were
formed in the interaction. 
Super star clusters 
are similar in compactness and luminosity to globular clusters when scaled to a 
common age.  
NGC 1569, for example, an Im galaxy that
is undergoing a burst of star formation, has made two super star clusters
in the recent starburst (see, for example, O'Connell, Gallagher, \& Hunter
1994), and many other super star clusters have been
found in galaxies that have undergone interactions.
In addition to learning about NGC 4449, locating additional super star clusters
would
be interesting in helping to determine what conditions are necessary for the 
formation of super star clusters, and by inference globular clusters.
Understanding the formation of globular clusters would help us better 
understand our own galaxy's evolution as well as that of NGC 4449.

When one looks at the visible images of NGC 4449, an object in the
center of the galaxy stands 
out because of both its brightness and its size.  Its brightness suggests 
that it could be one of these
super star clusters.  Alternatively, it has informally been referred to by one
of us as the nucleus
of NGC 4449, but since Im galaxies do not have nuclei, this would
be highly unusual. 
Here we are using the term ``nucleus'' to mean a centrally concentrated
substructure of the galaxy that has been an integral part of the galaxy
from the galaxy's formation.
One of our goals in this study was to determine the
nature of the central object. One important piece of information
is its age: a nucleus, by our definition, must be old; a star cluster
that just happened to form near the center need not be old.

To answer these questions, we used broad-band images obtained with
the Hubble Space Telescope ({\it HST}) to identify star clusters and determine
integrated photometry.
We compared the integrated photometry to star cluster 
evolutionary models to estimate ages.
To further set a limit on the
age of the central object, we obtained near-infrared spectra of the 
stellar $^{12}$CO(2,0)
and $^{12}$CO(3,1) absorption features.  The appearance of these features
signals the presence of red supergiant
(RSG) populations and hence can put a lower limit on the age of the cluster 
(Olofsson 1989).
From the integrated magnitudes and sizes 
of the clusters we determined the nature of the star
clusters, including
the central object, and compared them to other clusters.

\section{The Data}

\subsection{The {\it HST} Data}

NGC 4449 was imaged with the Wide Field and
Planetary Camera 2 (WFPC2) on {\it HST} on 1995 October 5, 1997 July 28,
and 1998 September 1, and we have obtained these data from the archives.
The WFPC2 consists of four mosaiced CCDs: the PC has a resolution
of 0.0456\arcsec\ and observations taken in 1998 placed
the center of the galaxy in the PC. 
The 1998 observations were offset from the position
of the 1997 observations presumably in order to cover more of the galaxy.
The 1995 observations placed the center
of the galaxy in WF2. The three WF CCDs have a resolution
of 0.0996\arcsec, which is 1.9 pc at the distance of the galaxy. 
In the 1995 observations 
exposures were taken through filter F606W.
The 1997 and 1998 observations 
included exposures through F170W, F336W, F555W, and F814W.
The F170W filter is a far-ultraviolet filter centered 
at $\sim$ 1700 \AA, and F606W is a red optical filter centered
$\sim$ 5900 \AA.  F336W, F555W, and F814W can be transformed to the
Johnsons/Cousins UVI filter system.
The footprints of the WFPC2 in these three sets of observations are shown
superposed on a ground-based image of NGC 4449 in Figure \ref{figfoot},
and the exposures are listed in Table \ref{tabhst}.
Multiple images of the same field through the same filter were combined
to remove cosmic rays.

We calibrated photometry of the clusters using
the synthetic zero points given by Holtzman \et\ (1995a) in
their Table 9. We also converted the F336W, F555W, and F814W
photometry to the Johnson/Cousins UVI system using formulas presented in
Holtzman \et\ The instrumental magnitudes
were corrected for reddening and the red leak in F336W, as discussed
below, before
converting to UVI, using the methods of Holtzman et al.

\subsection{Near-IR Spectroscopy}

We obtained longslit near-infrared spectra of the central object
in order to examine the stellar CO absorption features, which are an age
diagnostic because of their sensitivity to the presence of RSGs. 
The spectra were obtained on 1996 April 25 
with the Ohio State Infrared Imager/Spectrometer (OSIRIS) on the 1.8 m 
Perkins telescope.
They cover 21900 \AA\ to 23900 \AA\ at 8 \AA\ per pixel.
The cluster was observed in alternating positions on the slit about 10--20\arcsec\
apart.
One position was used as a sky observation for the other. 
The bright stars 
HR 4811 and
HR 4785, close in airmass to NGC 4449, were observed before and after blocks
of observation of NGC 4449 in order to remove absorption features due to 
the earth's atmosphere.

We corrected the data for the fact that the number of detected electrons
is a non-linear function of the incident photons 
using a series of observations of an illuminated 
white screen hanging in the dome.
The sequence consisted of a series of observations of increasing exposure times
interleaved with 1 second exposures; the 1 second exposures
were used  to remove drifts in the lamps' intensities.
Dark observations were subtracted from the data.
Regular dome flats were also taken and the data were flat fielded using these.
Sky frames were subtracted from object frames,
one-dimensional spectra were extracted, and galaxy spectra 
sandwiched between each two star observations
were combined.
Night sky lines in spectra with the sky not subtracted
were used to determine the wavelength scale and linearize
along the dispersion axis. The night sky lines were identified using a line list
provided by M.\ Hanson (private communication) which came originally
from C.\ Kulesa (private communication to M.\ Hanson).
The cluster spectra were divided by the nearest spectrum of either
HR 4811 or HR 4785
in order to remove the telluric absorption. 
The cluster spectra from the blocks between the star observations
were combined to produce a single, final cluster spectrum.
The continuum was fit and the spectrum divided by the fit.

\section{Data Analysis Issues}

\subsection{Reddening}

The foreground reddening has been estimated by Burstein \& Heiles (1984) to be
zero.  The internal reddening E(B$-$V) in HII regions in NGC 4449 is 
on average 0.1 magnitude (Hunter \& Gallagher 1997). The bulk 
of the stars, however, are not likely to be as 
reddened as HII regions. Therefore, 
we adopted an E(B$-$V) of 0.05 magnitude for correcting the broad-band
photometry for internal reddening.
We did not consider spatial varitions in E(B$-$V).

Holtzman \et\ (1995a) showed that the reddening correction for {\it HST}
filters is
a function of the spectrum of the object. Holtzman \et\ gives the
reddening corrections for an O6 and a K5 star in their Table 12.  
Since the spectral energy distributions of our clusters fall between these
two extremes and since the corrections are small, 
we have simply used the average of the correction for an
O6 and for a K5 star. Thus, we use extinction values of
A$_{336}$$=$0.24, A$_{555}$$=$0.16,
and A$_{814}$$=$0.10.
To correct the F336W filter, we first correct 
for its red leak, as described below.

\subsection{Red-Leak in F336W}

Because of the sizeable leak at red wavelengths, 
red photons can contribute significantly to the counts in
the F336W filter.
To determine the red leak as a function of the
observed F555W$-$F814W color, we used 
the STSDAS software to simulate the {\it HST} 
throughput for various filters and for different incident
blackbody spectra.  With an 
E(B$-$V)$=$ 0.05 and the LMC reddening curve, appropriate for a modestly
metal poor galaxy, we estimated the F336W magnitude through the actual
filter and through the filter with a cutoff.
The red leak was taken to be any flux contribution from $\geq$4000 \AA,
after the definition of Holtzman \et\ (1995b).
The correction to the F336W magnitude was 0.008 magnitude for an object with 
an F555W$-$F814W of 0.6 and
increases to 0.05 magnitude for colors above 1.2.
These corrections were applied to the integrated cluster magnitudes 
in F336W 
using the observed F555W$-$F814W cluster color.

\subsection{Distance to the Galaxy}

We assume a distance of 3.9 Mpc, determined
from a Hubble constant of 65 \kms\ Mpc$^{-1}$. 
Although this distance is uncertain, it is consistent with its
being a member of the CVnI group of galaxies (de Vaucouleurs 1975).
However, a distance of 5 Mpc has been used by some
(Sandage \& Tammann 1975, Aaronson \& Mould 1983).

\subsection{Metallicity}

The oxygen abundance (12$+$log(O/H)) of HII regions in 
NGC 4449 is 8.3 (Talent 1980), which
corresponds to a Z of about 0.005 or one-third solar. When we compare
photometry of the star clusters to cluster evolutionary models below
we will use both Z$=$0.004 and Z$=$0.008.
The $Z=0.004$
models are closer to the metallicity of NGC 4449,
but because of uncertainties in the models,
we looked at both sets and found that the Z$=$0.008 models fit
cluster colors better.
This was also the case in fitting the star clusters in the irregular
galaxy NGC 1569 (Hunter \et\ 2000).

\section{The Central Object}
  
\subsection{Age}

The near-infrared spectrum of the central object is shown in Figure
\ref{figco}.
We detected the $^{12}$CO(2,0) and $^{12}$CO(3,1)
stellar absorption features around 2.3 microns, and the
equivalent widths of these features are given in
Table \ref{tabco}. For comparison we include individual stars in 
the Table that we measured
as well as some observed
by Hunter \et\ (2000).
The equivalent widths are comparable to those of cool stars.
Thus, the central object contains a significant population of cool,
luminous stars. 

In a coeval stellar population,
prior to the onset
of the RSG phase, a young coeval cluster will exhibit
no CO features. Then as the massive stars in the cluster evolve to the RSG phase,
the CO features
appear in the integrated cluster spectrum.
For the metallicity of NGC 4449, this occurs
at $\sim$7 Myrs (Mayya 1997, Origlia \et\ 1999).
The red supergiant peak only lasts $\sim$5--7 Myrs
(Bica \et\ 1990, Origlia \et\ 1999), but after that
other cool stars continue to contribute to the CO feature, including
an AGB phase at $\sim$100 Myrs (Olofsson 1989, Bica \et\ 1990).
Thus, the CO features can be used to set an upper or lower limit
on the age of the clusters depending on whether the features are
present or not.
We conclude that the central object in NGC 4449 is older than 7 Myrs.

To further explore the age of the central object, we measured its
integrated UVI colors from the {\it HST} images. The structure of
the central object, however, 
presented a problem in deciding what aperture size to use.
The central object is a bright peak in an extended, higher surface brightness
distribution of stars near the center of the galaxy. So, where does
the object end? Is it just the core (radius 0.68\arcsec)
or does it include the larger
central region (radius 1.8\arcsec)?
We opted to consider the central object as
just the core of this central region.
We felt that it was more likely
that only the core was a gravitationally bound entity and, hence,
star cluster.
Therefore, we performed integrated photometry on the central object
of NGC 4449
using an aperture of 0.68\arcsec\ radius, and subtracting the surrounding
background galaxy by determining the mode in an annular region just
beyond that radius.

We compare the colors of the central object to
cluster evolutionary models prepared by Leitherer \et\ (1999)
in Figure \ref{figccd}.
We use the
instantaneous star formation models appropriate to a star cluster, although
we did examine the continuous star formation models.
We used the metallicity $Z=0.004$ and $Z=0.008$
models and a Salpeter (1955) stellar initial mass function.  
The Z$=$0.008 model, in fact, appears to better fit the colors of the central
object and other clusters
in NGC 4449. 

We see that the UVI colors of the central object are, in fact, those
of a star cluster of age 8--15 Myrs. This age range is consistent with
the age limit of $\geq$7 Myrs predicted by the appearance of the
stellar CO absorption features in the near-infrared spectrum of the object.
It also means that the central object, or cluster 1, as we will now refer to it,
is in the RSG phase, meaning that the light is dominated by evolved massive
stars entering this evolutionary phase.

\subsection{Comparison to Super Star Clusters}

From the {\it HST} data, we find that cluster 1 has a
half-light radius \rhalf\ of 5 pc and an integrated M$_V$ of $-$12.6.
How does this compare to properties of super star clusters?
The four young superstar clusters in the LMC, NGC 1569, and NGC 1705 have 
\rhalf\ of 1.7--3.4 pc 
(Hunter \et\ 1995; O'Connell \et\ 1994). 
Thus, cluster 1 is not as compact
as these.
Globular clusters have an average
\rhalf\ $\sim$3 pc, but with a range from 1--8 pc
(van den Bergh \et\ 1991).
Thus, the central
cluster is comparable in compactness to some globular clusters
but larger than average. It may be comparable to the luminous clusters
found in the interacting galaxies ``The Antennae'' by Whitmore \et\ (1999).
There the effective radii of many clusters that are luminous enough to
be globular-like are larger than the typical globular cluster.
Whitmore \et\ suggest that in some cases 
the larger radii may be due to the fact that outer stars have not
yet been stripped from the young clusters by the tidal actions of the host
galaxies as they have in the old Milky Way globulars.

The integrated M$_V$ of the central cluster
in NGC 4449 is
$-$12.6. The clusters in NGC 1569 and NGC 1705 are close enough in
age to cluster 1 to allow a direct comparison of M$_V$.
The two clusters in NGC 1569 have M$_V$ of $-$14.0 and $-$13.0,
and that in NGC 1705 has an M$_V$ of $-$11.1.
R136 in the LMC is much younger, 1--2 Myrs; the evolutionary models
predict that R136 would fade 0.3 magnitude by the time it reaches
10 Myrs of age. Thus, R136 would be expected to have an M$_V$ of
$-$10.8 when it is about the age of cluster 1. Therefore, we see
that cluster 1 is well within the range of luminosities of
the other super star clusters and is comparable to the brightest of these.

Using the Leitherer \et\ (1999) models, we estimated a mass for 
the cluster based on the
integrated brightness.  
In Figure \ref{figcmd} we plot the clusters, including cluster 1,
with the cluster evolutionary models. The models can be slid up and down
in M$_V$ to match any cluster and the mass is appropriately scaled from
the model mass of 10$^6$ M\solar. The models in Figure \ref{figcmd}
have been dimmed 1.7 magnitudes in M$_V$ in order to match cluster 1.
The central cluster, thus, has a mass of 
2$\times$10$^5$ M\solar. This mass is similar to that of the smaller 
globular clusters which have masses of 10$^4$--10$^6$ M \solar\
(Pryor \& Meylan 1993).

Therefore, all indications are that the central cluster in NGC 4449 
deserves to be considered a super star cluster. It is comparable to 
what a small globular cluster was probably like when globulars were only
about 10 Myrs old if stars are present in normal proportions over the
full range of stellar masses.

\subsection{Surface Photometry}

Because the central star cluster is relatively 
well resolved in the {\it HST} images, 
we have examined surface photometry of it. The results are
shown in Figure \ref{figmu}. Photometric apertures ranged from 3 to 15 pixels
in steps of 3 pixels. The choice of 3 pixels, which corresponds to
0.14\arcsec, was dictated by the
desire to exceed the point-spread-function profile in the PC.

We have fit the V-band surface brightness profile of cluster 1
with a King model (King 1962) appropriate for globular clusters.
We explored the parameter space of reasonable solutions, determining
the best fit and the range of values that are reasonable for the three
variables. The uncertainties are large and 
are determined from the range in reasonable fits. The best fit 
is shown as the solid line in Figure \ref{figmu}. The fit uses
a $\mu_0$ of 13.5$\pm$0.3 magnitudes arcsec$^{-2}$ and
a core radius r$_c$ of 1.9$\pm$0.3 pc. The concentration index,
$c=\log (r_t/r_c)$ where $r_t$ is the terminal radius, is roughly 1.5. 
An average core radius for globular clusters is
2.6$\pm$3.9 pc with a total range of 0.2--23 pc
(Djorgovski 1993; Trager, Djorgovski, \& King 1993).
The core radii of one of the super star clusters in NGC 1569 is
0.6$\pm$0.1 pc (Hunter \et\ 2000).
For Milky Way globular clusters the average concentration index
is 1.4$\pm$0.4, with a total range of 0.6--2.4.
The concentration index of the clusters in NGC 1569 are
1.7--2.0.  Thus, we see that cluster 1 in NGC 4449 has structural
parameters comparable to other super star clusters and globular
clusters.

In terms of radial trends, we see that cluster 1 appears to get
redder by $\sim$0.15 magnitude
in both V$-$I and U$-$V with increasing radius in the inner
0.4\arcsec. Beyond that the cluster gets bluer again.
Some of this radial color variation could be due to contamination of
the broad-band images by emission-lines. The region is known to
contain nebular emission (see, for example, Hunter 1982;
Hunter, Hawley, \& Gallagher 1993). However, the spatial resolution of
these ground-based images is too coarse compared to the {\it HST}
broad-band images to be able to determine
the contamination of the cluster photometry. Since
the emission-lines would probably affect V more than I, their
presence would make V$-$I bluer.

\subsection{Structure}

Up to this point, the properties of the central cluster seem to point
to its being a normal super star cluster. However, closer inspection
of the {\it HST} images reveals that this picture may be too simple.
A shallow inspection of the cluster to look for very bright features,
reveals
what looks at first glance like a tiny barred spiral
galaxy seen edge on at the center of the cluster.  
A contour plot of the cluster is shown in
Figure \ref{figcont}. It 
shows a bright condensation 0.1\arcsec\ (1.9 pc) in diameter sitting in the
middle of an elongated, bar-like structure that is
0.5\arcsec\ (9.5 pc) long.
The minor to major axis ratio of the structure is 
0.6, and the cluster itself is essentially round in the outer parts. 
In the rest of the {\it HST} field of view of
the galaxy, there are no other images that look like 
this: all candidate background galaxies are spheroidal in appearance.
It would be very unfortunate indeed if the
only background disk galaxy occurred right behind the center of the central
star cluster.  Consequently, we assume that what we see is not a background 
galaxy, but rather structure in the cluster itself. Such structure is
very unusual for a star cluster.

\subsection{Nucleus, Star Cluster, or Cannabilism?}

The evidence suggests that the central object in NGC 4449 is only $\sim$10 Myrs
old. It is, therefore, not a nucleus of the galaxy in the traditional sense.
Instead we have suggested that the central object is a super star cluster.
But, given the tiny nucleated-bar structure in the center of the cluster discussed
in the previous section, we would like to offer another twist to the interpretation:
that this structure is the result of the interaction scenario that
also produced the counter-rotating gas systems and extended
gas filaments and streamers.

Models of a larger galaxy swallowing a smaller galaxy can produce
counter-rotating cores and, for some mass ratios, they can produce
a core within a core (Balcells \& Quinn 1990). Furthermore,
the most tightly bound debris can become a distinct subsystem
in the nucleus of the disk (Quinn, Hernquist, \& Fullagar 1993)
and produce unusual nuclear features (Hernquist 1991).
Alternatively, the interaction could have helped to channel gas to the
center of NGC 4449 (Noguchi 1988).
The in-falling gas could then account for the
large complexes of gas
forming a partial ring around the optical galaxy.
In addition in-falling debris could have resulted in peculiar
structures at the center of the galaxy.

We suggest that the peculiar structure at the center of cluster 1 in 
NGC 4449 is debris from cannibalism
of a smaller companion that has fallen to the center of the galaxy.
However, exactly how the bar structure and the enveloping star cluster fit together
is not clear.  In this scenario the material composing cluster 1 could
have been stars accreted from the companion or gas that was accreted
and then formed stars.
Since the integrated colors of the cluster suggest a young age, 
in situ star formation seems more plausible.
The appearance of the structure at the center of cluster 1
also suggests that it is a distinct sub-system that
could be rotating in a different plane than the rest of the galaxy.
The HI gas already suggests that there are other, larger, sub-systems at
very different inclination angles to the optical galaxy (Hunter \et\ 1999).

\section{Other Star Clusters}

\subsection{Identification of Clusters}

We identified additional star clusters on the basis of
their visual appearance in the F555W image.  We looked for bright, compact
clusters that are resolved compared to an isolated star. 
Radial profiles yield FWHMs of the clusters, and we also performed integrated
photometry.
The limits of each of the clusters were taken as the radius where the
cluster visually
merged into the background, usually twice the FWHM.
The contributions from the sky and background galaxy were measured as the mode
in an annular region just beyond
the cluster aperture.
We list the star clusters and their integrated photometry in
Table \ref{tabcl}.
The uncertainties are determined from Poisson statistics and read noise.
The central cluster of the galaxy is 
labelled cluster 1 and the rest follow numerically in order of
Right Ascension.
In Table \ref{tabcl} we include the distance to the cluster 
in the plane of the galaxy using the center of the galaxy determined
from the symmetry of the outer isophotes of the optical galaxy (Hunter,
van Woerden, \& Gallagher 1999).

\subsection{Ages and Distribution}

We show the integrated colors and magnitudes in color-color and
color-magnitude diagrams in Figures \ref{figccd} and \ref{figcmd}.
Cluster colors range from (V$-$I)$_0$ of $-$0.3 to 1.05 and
(U$-$V)$_0$ of $-$1.2 to 0.9. We interpret the range in colors
as due to a spread in ages of the clusters although metallicity must
also be a parameter if age is.
We see then that the clusters span the full age range of the models, from
about 4 Myrs 
to 1 Gyr. Furthermore, 
there does not appear to have been an unusual peak in cluster production within
the past 1 Gyr that we
can use to help date an unusual event in the life of the galaxy.

In addition, there is no pattern in the spatial distribution of the clusters with
respect to age.  There tend to be younger clusters near regions that have
previously been identified as star-forming regions -- which is not unusual.

The exception is a group of 14 very red objects---seen in the lower, right of
the plots in Figure \ref{figccd}, (V$-$I)$_0$$>$0.8 and (U$-$V)$_0$$>$0.2,
including two that are just beyond the Figure boundaries.
Two of these objects, clusters number 30 and 40, lie near the center of the
galaxy, but the rest lie in the outer parts of the galaxy.
These objects are redder in (V$-$I)$_0$ than the models. They also
lie in a region occupied by early-type galaxies (Poulain 1988)
and some globular clusters (Reed 1985).
They are
also all round in shape. Therefore, while these objects could
be bonifide old globular clusters in NGC 4449, we 
are suspicious that they are background elliptical galaxies.
Without further data, we cannot tell whether these
are galaxies or star clusters, and so we will not include them 
in the discussion that follows.

\subsection{The Nature of the Clusters}

To learn more about the clusters, we have corrected the observed
M$_V$ to the M$_V$ the cluster would have had at an age of 10 Myrs.
We used the colors obtained from our cluster photometry and comparison
to the cluster evolution models to determine a rough age and luminosity
correction for each cluster.  
A histogram of the M$_V$ is shown
in Figure \ref{figMv}. The median M$_V$ is $-$9.

Figure \ref{figfwhm} is a
histogram of the FWHM's of the clusters in the F555W filter.
The FWHM range from 1 to 9 pc, with a median of 4 pc.
For comparison, NGC 1569's super star clusters
have FWHM's of 2.6 and 2.3 pc.
Thus, some of the clusters are as compact as globulars or super star
clusters, but others are not.

Three clusters have M$_V$ corrected to 10 Myrs that are quite luminous,
$\leq-11$. In addition they are fairly compact, FWHM$\leq$4.5 pc.
These characteristics place them 
in the realm of super star clusters. One of these is cluster 31,
whose colors yield an age of about 20 Myrs. It is located just 21 pc
from cluster 1, and is only a little bit older than cluster 1.
The other two super star cluster candidates,
clusters 27 and 47, have colors that correspond to an age of 0.75--1 Gyr.
Cluster 27 is located fairly near clusters 1 and 31, while cluster 47
is located to the north near the regions of recent star formation.

There are also 14 star clusters with M$_V$ between $-$10 and $-$11.
These clusters could be comparable to the populous clusters found in
the LMC. They are intermediate between open clusters and globular
clusters in terms of size and mass. One young populous cluster in the
LMC, NGC 1818, for example, has an M$_V$ corrected to an age of 10 Myrs 
of $-$10, and its \rhalf\ is 3.2 pc (Hunter \et\ 1997).
The populous-like star clusters in NGC 4449 span ages from 10 Myrs to
0.75--1 Gyr.

\section{Conclusions}

We have examined the nature of the central object in the irregular galaxy
NGC 4449 through a near-infrared spectrum of stellar CO features
and broad-band {\it HST} images.
The central object in NGC 4449 has many of the characteristics of
a super star cluster. The broad-band colors and near-infrared spectrum
are consistent with an age of 8--15 Myrs and with the cluster being
in the phase dominated by RSGs.
While not as compact as many super star clusters, it is still within
the realm of globular cluster radii. In addition it is as luminous
as many super star clusters. However, there is a peculiar nucleated-bar
structure at its center that is not seen in star clusters. We suggest
that this structure is related to the interaction that has produced
the counter-rotating gas systems and extended gas streamers.

We have used the {\it HST} images to identify additional compact star
clusters.
We identified 61 cluster candidates in NGC 4449 but
suspect that 14 of these may be background elliptical galaxies.
The cluster ages deduced from integrated photometry and compared to
cluster evolutinary models span the timescale 
of the models up to 1 Gyr.  Apparently, NGC 4449 has been
an actively star-forming galaxy for quite some time.
Therefore, we did not find a ``special time'' when a large number of the 
star clusters in NGC 4449 formed, and so 
we cannot connect cluster formation to
the date of an interaction.  
There are three other clusters that may be comparable to super star clusters
and many more that may be comparable to the populous clusters found
in the LMC. The rest are more modest groupings of stars.
The formation of the more massive clusters, but not necessarily the
super star clusters, may simply reflect the 
high level of star formation activity in NGC 4449 rather than its
past interaction since NGC 4449 falls on the relationship found
by Larsen \& Richtler (2000) between the integrated properties of
galaxies and their young massive star cluster content.

\acknowledgments

The Ohio State University 
provided access to their OSIRIS spectrograph when it was on the Perkins
1.8 m telescope at Lowell Observatory, and we would like to thank
Mark Wagner for his efforts to support OSIRIS.  AEG would like to thank the
National Science Foundation for providing funding for the Research Experience
for Undergraduates under grant number 9988007 to Northern Arizona University
and Kathy Eastwood for directing the program. DAH is grateful for funding
from grant AST-9616940 from the National Science Foundation.

\clearpage

\begin{deluxetable}{rrrrc}
\tablecaption{{\it HST} Images\label{tabhst}}
\tablewidth{0pt}
\tablehead{
\colhead{} & \colhead{} & \colhead{ORIENT\tablenotemark{c}}
& \colhead{} & \colhead{} \nl
\colhead{RA (2000)\tablenotemark{a}} 
& \colhead{DEC (2000)\tablenotemark{b}} & \colhead{(degree)} 
& \colhead{Filter} & \colhead{Exposures}
}
\startdata
12 28 12.9 & 44 05 43.2 & $-$175.74 & F606W & 2$\times$80 s \nl
12 28 11.5 & 44 05 41.0 & $-$51.10  & F170W & 2$\times$400 s \nl
           &            &           & F336W & 2$\times$260 s \nl
           &            &           & F555W & 1$\times$200 s \nl
           &            &           & F814W & 1$\times$200 s \nl
12 28 09.6 & 44 05 12.0 &  127.82   & F170W & 2$\times$400 s \nl
           &            &           & F336W & 2$\times$260 s \nl
           &            &           & F555W & 1$\times$200 s \nl
           &            &           & F814W & 1$\times$200 s \nl
\enddata
\tablenotetext{a}{RA units are hours, minutes, seconds.}
\tablenotetext{b}{DEC units are degrees, arminutes, arcseconds.}
\tablenotetext{c}{Image header keyword that can be
used to determine the position angle of North in the images.}
\end{deluxetable}

\clearpage

\begin{deluxetable}{lccc}
\tablecaption{Stellar CO Equivalent Widths \label{tabco}}
\tablewidth{0pt}
\tablehead{
\colhead{} & \colhead{} & \colhead{$^{12}$CO(2,0)}
& \colhead{$^{12}$CO(3,1)} \nl
\colhead{Object\tablenotemark{a}} 
& \colhead{Spectral Type} & \colhead{(\AA)} & \colhead{(\AA)} 
}
\startdata
Cluster 1     &        &  20  & 10   \nl
HR 4869       & A2V    &   0  &  0   \nl
HR 4811       & A5V    &   0  &  0   \nl
HR 4785       & G0V    &   0  &  0   \nl
HR 4883       & G0III  &   0  &  0   \nl
HR 4815       & G8     &   6  &  6   \nl
HR 4932       & G9II-III & 6  &  6   \nl
$\xi$And      & KOIII  &   9  &  8   \nl
35$\sigma$Per & K3III  &  21  & 18   \nl
HR 8726       & K5Ib   & 34   & 22   \nl
$\chi$Peg     & M2III  & 27   & 18   \nl
$\mu$Cep      & M2Ia   & 41   & 26   \nl
\enddata
\tablenotetext{a}{K and M stars are from Hunter \et\ (2000).}
\end{deluxetable}

\clearpage

\begin{table}
\dummytable\label{tabcl}
\end{table}

\clearpage

\figcaption{Footprint of the WFPC2 images for
each of the three sets of {\it HST} observations
of NGC 4449 shown superposed on a V-band ground-based image.
The ground-based image was obtained for us by P.\ Massey
using the 4 m telescope at Kitt Peak National Observatory
(Hunter, van Woerden, \& Gallagher 1999).
\label{figfoot}} 

\figcaption{Near-infrared spectrum of cluster 1. The
$^{12}$CO(2,0) at 22935 \AA\ and $^{12}$CO(3,1) at 23227 \AA\
stellar absorption features
are marked. The absorption line to the red of these is also a CO
line but we did not measure it.
\label{figco}}

\figcaption{Star clusters are shown in a UVI color-color diagram.
The solid curve in the upper panel is an evolutionary track for a cluster with
instantaneous star formation and a metallicity of 0.004 and
a Salpeter (1955) stellar initial mass function with an upper
limit of 100 M\protect\solar\ (Leitherer \et\ 1999). The
solid curve in the lower panel
are their models for a metallicity of 0.008.
Ages 1--9 Myrs in steps of 1 Myrs are marked with x's along these lines;
ages 10, 20, and 30 Myrs are marked with open circles.
The evolutionary tracks end at 1 Gyr.
The arrow in the lower left corner of the upper panel is a reddening line for
a change of 0.2 in E(B$-$V)$_t$. It represents the average of an O6 and
a K5 type spectrum with A$_V$/E(B$-$V)$=$3.1 and a Cardelli \et\
(1989) reddening curve.
Cluster 1 is plotted as the number ``1'' rather than a point in order
to identify it.
\label{figccd}}

\figcaption{Star clusters shown in UVI color-magnitude diagrams.
The solid curve is an evolutionary track for a cluster with
instantaneous star formation and a metallicity of 0.004 and
a Salpeter (1955) stellar initial mass function with an upper
limit of 100 M\protect\solar\ (Leitherer \et\ 1999); the dashed
line are their models for a metallicity of 0.008.
Ages 1--9 Myrs in steps of 1 Myrs are marked with x's along these lines;
ages 10, 20, and 30 Myrs are marked with open circles.
The evolutionary tracks end at 1 Gyr.
The evolutionary tracks have been dimmed 1.7 magnitudes in M$_V$
in order to match cluster 1;
for clusters of other masses the lines would slide up or
down in the diagrams.
\label{figcmd}}

\figcaption{Integrated and surface brightness profiles of cluster 1 in
M$_V$ and integrated and annuli color profiles of (V$-$I)$_0$
and (U$-$V)$_0$.
The solid curve in the surface brightness plot is
a King model fit. Quantities are corrected for reddening.
\label{figmu}}

\figcaption{F555W contour plot of cluster 1 and immediate surroundings
in NGC 4449. Every 10 pixels (0.4555\arcsec) are numbered along the 
perimeter.
\label{figcont}}


\figcaption{Histogram of the star clusters of NGC 4449
showing the frequency of M$_V$ corrected to an age of 10 Myrs using
the Leitherer \et\ (1999) cluster evolution model for Z$=$0.008.
\label{figMv}}

\figcaption{Histogram of the star clusters of NGC 4449 showing the
frequency of the FWHM's.
\label{figfwhm}}

\end{document}